**Ostwald's Rule of Stages in one-dimension**

Jiajun Chen[1,2†], Ying Xia[1,2†], Mingyi Zhang[2,3], Yu Huang[4,5], and James J. De Yoreo[1,2*]

[1]Department of Materials Science and Engineering, University of Washington, Seattle, WA 98195, USA

[2]Physical Sciences Division, Pacific Northwest National Laboratory, Richland, WA 99352, USA

[3]School of Aerospace and Mechanical Engineering, University of Oklahoma, Norman, OK 73069, USA

[4]Department of Materials Science and Engineering, University of California, Los Angeles, CA 90095, USA

[5]California NanoSystems Institute, University of California, Los Angeles, CA 90095, USA.

[†]Co-first authors

[*]Corresponding author. Email: james.deyoreo@pnnl.gov

**Abstract**

Ostwald's Rule of Stages, which is one of the most widely observed phenomena associated with crystallization of polymorphs, follows naturally from the thermodynamics of nucleation. However, most observations of its manifestations have been limited to three-dimensional crystals and its validity in one-dimension, where no nucleation barrier exists, remains unclear. Here we investigate the two-dimensional assemblies and phase transformation mechanisms of a peptide that forms two distinct phases on graphite via one-dimensional nucleation using in situ atomic force microscopy. We find that the evolution of phases illustrates Ostwald's Rule, but does so for purely kinetic reasons, and that the stable phase replaces the metastable via a dissolution-reprecipitation mechanism enabled by inherent fluctuations of the phase boundary. The findings provide general insights into the growth and transformation mechanisms of coexisting two-dimensional phases and thus delineate a strategy for capturing transient two-dimensional structures.

**Introduction**

Ostwald's Rule of Stages is one of the most widely observed phenomena associated with crystallization in polymorphic systems[1–6]. The signature feature of this empirical rule first proposed by Ostwald in 1897 is that less stable phases appear first and are sequentially replaced by the more stable phases in order of increasing stability[1,2,7–10]. Though there is no a priori reason for Ostwald's Rule to be universally true[2,9,11], a thermodynamic rationale for the rule can be found in the inherent scaling of the free energy barrier for nucleation with a crystal's surface energy[12]: Because metastable phases are likely to be more structurally similar to their precursors in the growth media (e.g., a solution or melt), polymorphs typically exhibit a trend towards increasing surface energy as the stability of the crystal phase increases[13]. Thus, the formation of more stable phases is opposed by larger barriers and, hence, the time required for their nucleation is longer[14]. Once a more stable phase forms, however, the saturation state of the media — the solute concentration in the case of a solution — decreases towards the equilibrium value with respect to the more stable phase, driving the disappearance of the less stable ones and resulting in a scenario often termed "dissolution-reprecipitation."

The above argument on Ostwald's Rule, which is closely associated with free energy barrier for nucleation, makes sense for two- and three-dimensional (2D and 3D) crystals because the difference in the dimensionality of the bulk and the interface — volume vs. surface area in three-dimension and area vs. edge length in two-dimension — is what gives rise to the free energy barrier to nucleation[12]. In contrast, for one-dimensional (1D) systems no such difference in dimensionality exists and thus there is no free energy barrier available to enforce Ostwald's Rule[12,15].

Previously, we showed that, even for 2D crystals, nucleation may follow the physics of 1D nucleation provided there is high anisotropy in the intermolecular bonding along the two orthogonal directions of the 2D lattice so that the crystals nucleate one row at a time[15]. This observation raises the question of whether Ostwald's Rule still holds for 1D and quasi-1D systems even though there is no free energy barrier and nucleation is controlled by the kinetics of the individual attachment events that create the first structural unit, rather than the thermodynamics of an ensemble of such units. Peptides are inherently anisotropic in their binding and are well-known to assemble into 2D crystalline structures on surfaces. Moreover, numerous peptide and protein systems are known to exhibit metastable phases in both two-dimensions[16–21] and three-dimensions[22–25]. Thus, peptides that assemble into 2D crystals on surfaces provide a good platform upon which to answer the question of whether Ostwald's Rule of Stages is observed in 1D and quasi-1D crystal systems and, if so, why.

To address this question, we used in situ atomic force microscopy (AFM) to explore the assembly and transformation of 2D crystals formed from the peptide MoSBP1 (Tyr-Ser-Ala-Thr-Phe-Thr-Tyr; YSATFTY) on highly oriented pyrolytic graphite (HOPG) in aqueous solution. MoSBP1 assembles into ordered 2D structures on $MoS_2$ surfaces following a row-by-row growth pathway[15]. It can form nearly identical 2D crystals on HOPG, however, in contrast to the rows formed on MoS$_2$, on HOPG the rows exhibit high mobility that allows for more complex, aggregation-based assembly pathways. Moreover, as we show here, at sufficiently high supersaturation, two phases with distinct epitaxial relationships emerge on HOPG. The metastable (M) -phase forms more rapidly than the stable (S) -phase, but it is eventually replaced by the latter through a dissolution-reprecipitation mechanism. The details of that process reveal a version of Ostwald's Rule of Stages for 1D systems based purely on kinetic considerations, as well as the crucial role of M-phase fluctuations in enabling the S-phase to nucleate and eventually replace the M-phase. General trends in chemical kinetics imply that, just as the thermodynamic rationale for Ostwald's Rule is widely applicable to 2D and 3D systems, this kinetic rationale should be widely applicable to 1D and quasi-1D systems.

## Results

**Observation of a transition between two distinct coexisting phases**

To first determine what structures MoSBP1 forms on HOPG substrates, freshly cleaved substrates were incubated at room temperature with aqueous MoSBP1 solutions for a range of peptide concentrations. At low concentrations (~1 µM), only one phase —which we name S-phase for reasons explained below — appears during the early stage of assembly (Fig. 1a). The elongated 2D islands are mainly aligned along 3 crystallographically equivalent directions of the graphite lattice. The waviness of the islands reflects their mobility on the HOPG surface at this early stage, when the islands consist of a few rows or fewer. At higher concentrations (5 µM), initially two distinct assembled phases were observed to coexist (Fig. 1b). In addition to the S-phase with a height of ~0.8 nm (circled in black), which appears exclusively in low concentration solution (Fig. 1a), a shorter phase — which we name M-phase for reasons explained below — was observed having a height of only ~0.6 nm (circled in green and exhibiting a darker contrast in the AFM images). After approximately one hour, the shorter M-phase gradually disappears, and the entire surface becomes covered increasingly by the taller S-phase (Fig. 1c). This transition indicates that the shorter phase is a metastable state, thus named the M-phase, whereas the taller phase is the stable state, thus named the S-phase.

The *in-situ* AFM observation of the phase transition process (Fig. 1d-f) reveals the underlying nature of the transition. The white and green dashed circles in Fig. 1d highlight two representative regions where the transition from the shorter M-phase to the taller S-phase occurs. Over time, the area occupied by the M-phase gradually shrinks and is replaced by the S-phase. Notably, no lateral molecular movement was observed on the surface during this process, suggesting that the transition proceeds via dissolution of the M-phase followed by reprecipitation of the S-phase, a mechanism consistent with Ostwald ripening.

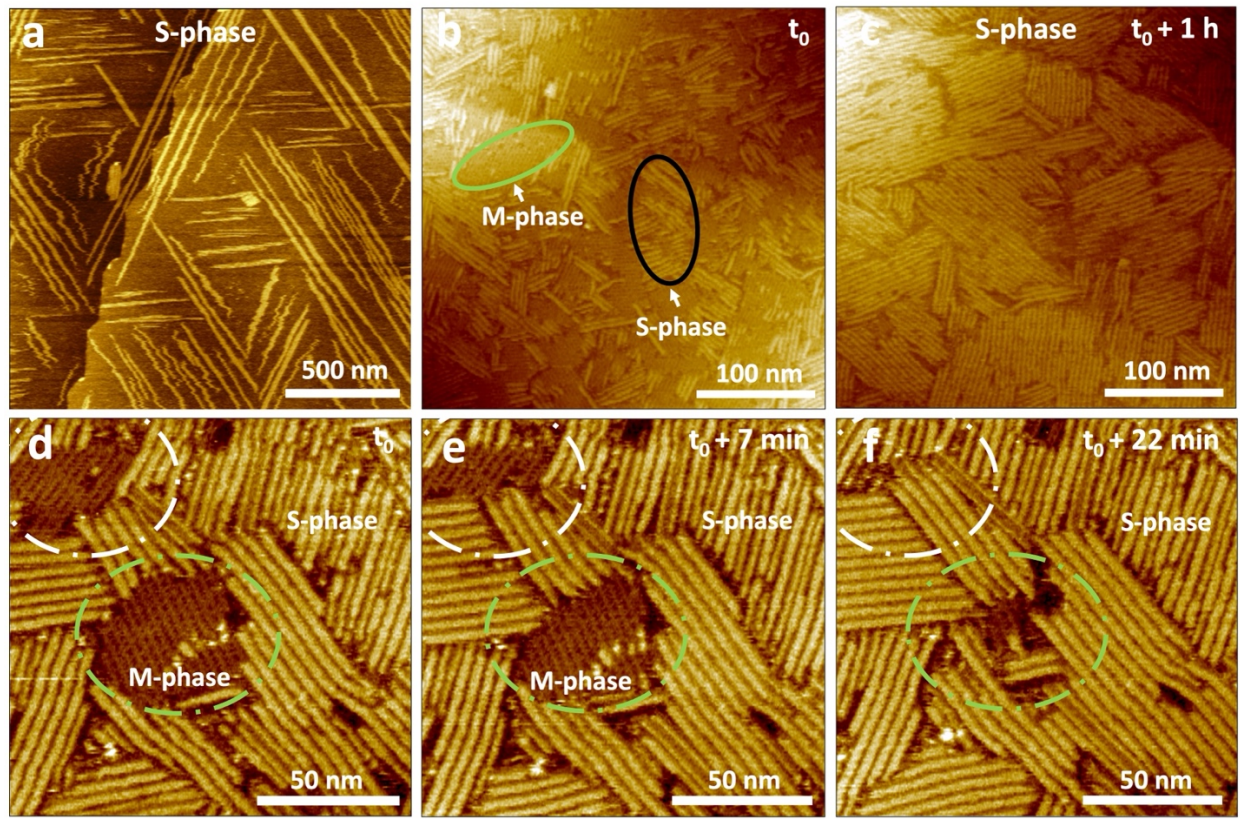

**Fig. 1| In situ AFM images of MoSBP1 on HOPG. a.** 1 μM of MoSBP1 on HOPG forms a single phase (S-phase) of elongated 2D islands for which waviness resulting from high island mobility is seen at the early stage of growth when the island consists of just a few rows or less. **b.** Initial assembly pattern MoSBP1 peptides on HOPG at a higher peptide concentration (5 μM), showing the coexistence of two distinct phases. Green and black circles represent the tall (S-phase) and short (M-phase) phases, respectively. The S-phase 1 is 0.2 nm taller than the M-phase 2. Height range: ~1.5 nm. **c.** Assembly pattern after one hour showing that the majority of the surface is now covered by the S-phase. **d-f.** *In-situ* AFM capturing the transition from the M- to S-phase at (**d**) $t_0$, (**e**) $t_0 + 7$ min, (**f**) $t_0 + 22$ min on HOPG. Height range: ~1.5 nm.

**Detailed structures of two distinct phases**

In addition to differences in height, the M- and S-phases exhibited distinct molecular arrangements within the constituent rows. S-phase islands (Fig. 2a), presented a structure similar to that we observed previously on $MoS_2$ (Supplementary Fig. 1): The islands were ~0.8 nm in height, indicating that they comprised monolayer-thick films, and consisted of parallel rows with a periodicity of ~4.5 nm (Fig. 2b,c, Supplementary Fig. 2). Molecular-resolution imaging showed that each row consisted of ~4.4 nm × 1.1 nm units, demonstrating the highly ordered structure of each row (Fig. 2c, Supplementary Fig. 3). However, the length of these building units on HOPG is smaller than that on $MoS_2$ where they are 4.7 nm in length, presumably due to the smaller lattice constant of HOPG compared to that of $MoS_2$. As was the case on $MoS_2$, the dominant row directions lay along the armchair directions of the graphite lattice, revealing an epitaxial relationship[15]. However, while rows or islands of MoSBP1 aligned along any other direction were not stable on $MoS_2$ and would gradually dissolve away[15], on HOPG, MoSBP1 islands lying along two other sets of three equivalent directions each were also observed (Fig. 2a), with the differences between these two sets of directions and the dominant armchair directions being 23.5° and -18.3°, respectively.

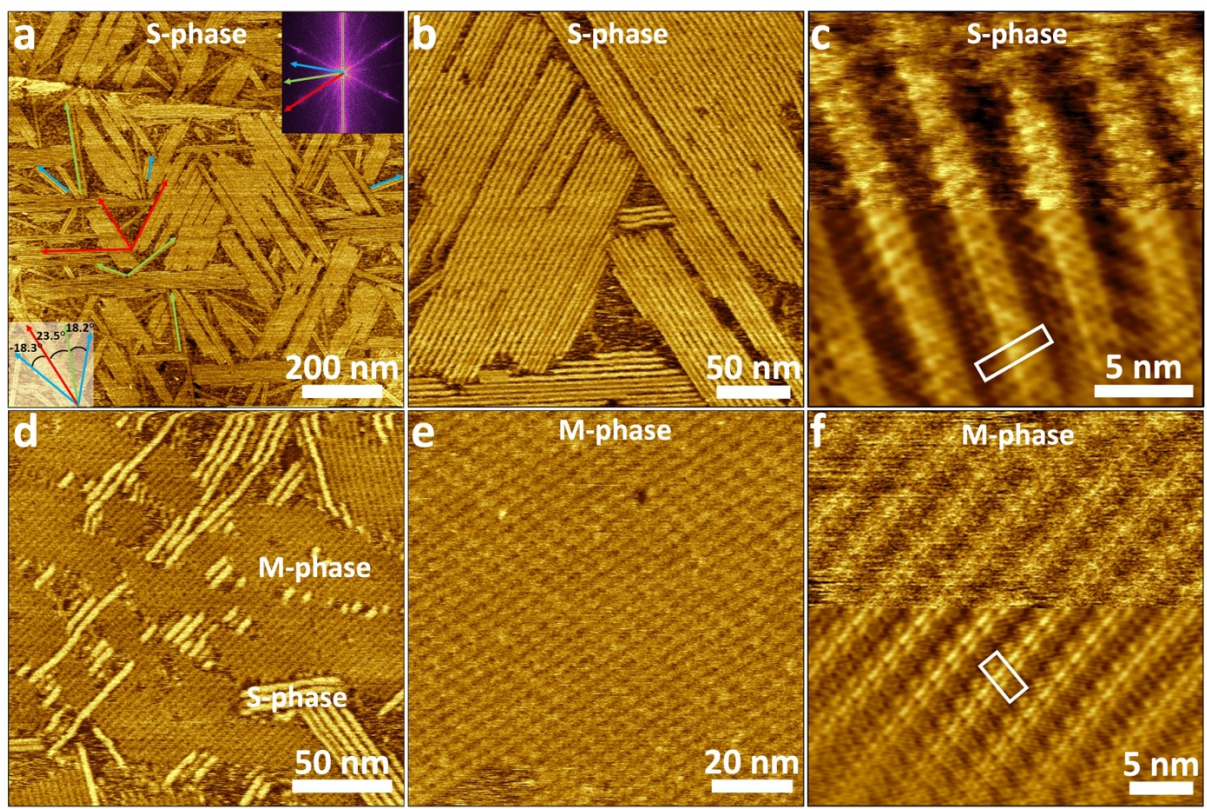

**Fig. 2| Detailed structures of MoSBP1 assemblies on HOPG. a**, S-phase after ~19 h of incubation at 5 µM. Inset shows the FFT of (a). Arrows delineate 3 different sets of directions; each containing 3 equivalent orientations. **b,c**, high-resolution images of the islands seen in (a). **d**, M-phase observed early in the assembly process at 5 µM, coexisting with S-phase, exhibiting a lower height (reduced brightness) than that of the S-phase. **e,f**, high-resolution images showing the detailed structure of the M-phase. The bottom halves of (c) and (f) are FFT filtered.

The M-phase that forms in the early stage of the assembly process on HOPG at MoSBP1concentrations > 4 µM (Fig. 2d-f, Supplementary Fig. 4) was observed to align only along the three equivalent armchair directions of the graphite lattice (Supplementary Fig. 5). The islands were also highly ordered and consisted of parallel rows with a uniform spacing of ~3.9 nm (Fig. 2e,f, highlighted in the white square) and heights of only ~0.6 nm, which is slightly smaller than that of the dominant S- phase (Supplementary Fig. 4). High resolution AFM imaging revealed that the rows consisted of repeating units with a size of 3.9 nm by 1.9 nm (Fig. 2f, highlighted in the white square). Comparing the unit size with the fully extended length of MoSBP1 (~2.5 nm), we conclude that these building units consist of peptide oligomers and are most likely dimers, as is the case both for the S-phase and for MoSBP1 on $MoS_2$ surfaces[15].

**Development pathways of S-phase and M-phase**

To understand the source of the observed coarsening process, we investigated the assembly and transformation dynamics of the M- and S-phases. By continuously monitoring assembly on HOPG (at 4 µM), we found that the M-phase islands were kinetically preferred, forming much faster than the S-phase islands, with the former appearing within 6 minutes after solution injection, but the latter requiring ~31 min of incubation time before formation in the region we monitored (Fig. 3). However, under no-flow conditions, the M-phase islands gradually dissolved as the S-phase islands grew, driven by the consumption of free monomers in solution. (Fig. 3, Supplementary Fig. 6, Supplementary Fig. 7). Thus, even at a higher peptide concentration of 5 µM, only the S-phase was observed after a sufficiently long incubation time (~19 h) (Fig. 2a).

Growth of the M-phase followed a row-by-row growth mechanism (Fig. 4), which was detailed in our previous study[15] of MoSBP1 assembly on $MoS_2$. These 2D crystalline arrays exhibited a classical nucleation and growth pathway, in that no transition from an amorphous precursor to an ordered structure was observed. In other words, M-phase islands formed directly through monomer-by-monomer addition to an ordered peptide cluster. The development of M-phase on HOPG was anisotropic, following the row-by-row growth pathway first seen on $MoS_2$, for which the longitudinal growth rate is linear in peptide concentration while the lateral growth rate is quadratic[15], because lateral growth requires nucleation of a new dimeric unit adjacent to an existing row (Fig. 4d-f), while longitudinal growth merely depends on one-dimensional (1D) extension via monomer attachment to the ends of the rows. Consequently, longitudinal growth was faster than lateral growth at small concentrations, leading to islands with high aspect ratios, but lateral growth took over at higher concentrations, leading to islands with small aspect ratios (Fig. 3, 4).

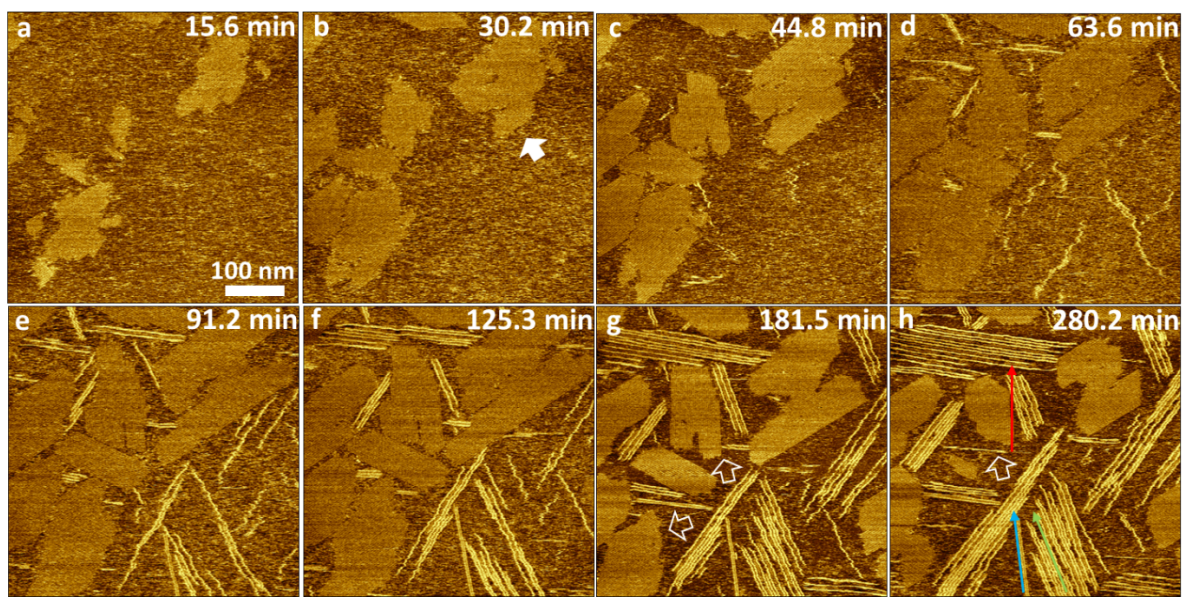

**Fig. 3| Snapshots showing the developments of the M-phase and S-phase at 4 µM MoSBP1 on HOPG.** M-phase islands appeared first and then gradually dissolved as the S-phase formed, slowly at first, but dominating over time. Arrows mark 3 different sets of directions of peptide rows. Solid and open white arrows mark the growth and dissolution of the M-phase, respectively. Colored arrows in (h) denote the 3 different sets of directions of the peptide rows.

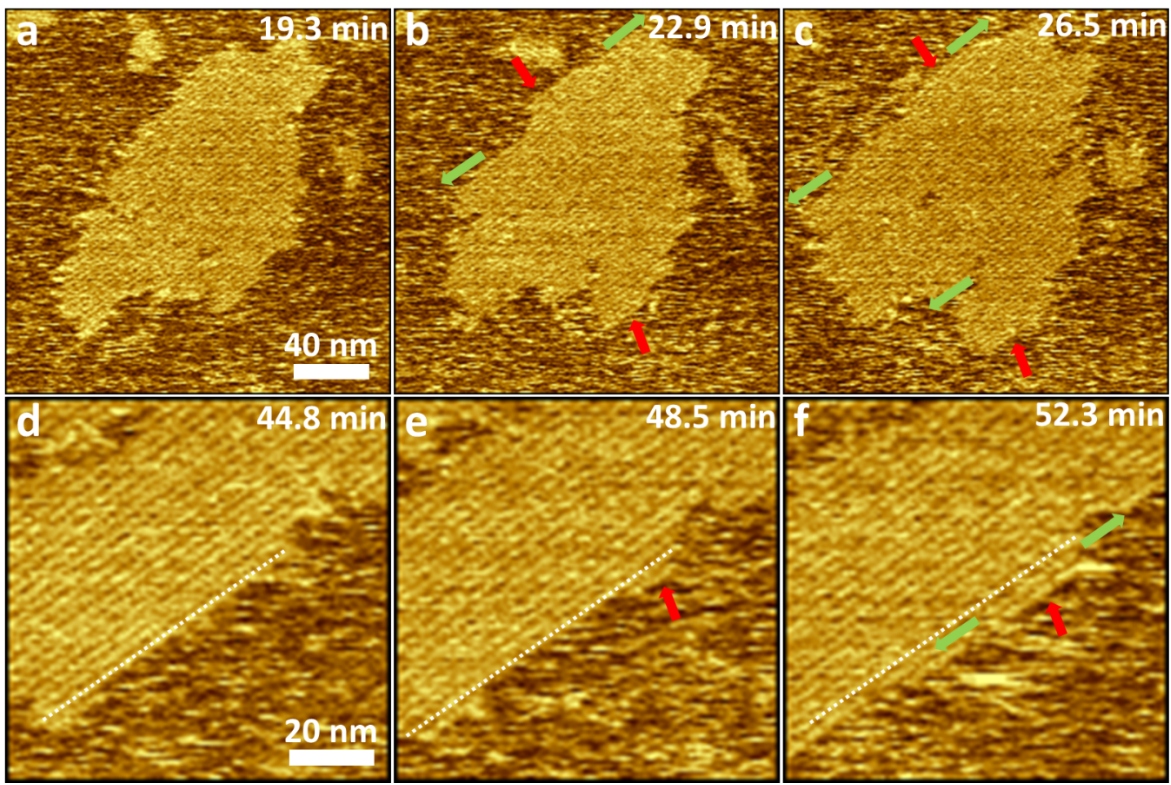

**Fig. 4| Row-by-row growth mechanism of M-phase (4 µM)**. **a-c**, longitudinal and lateral growths of a M-phase island. **d-f**, formation of a new row adjacent to an existing one (marked with a dotted line). Red arrows show the creation of a new row (lateral growth) while green arrows show the elongation of an existing row (longitudinal growth).

At an intermediate concentration of 4µM, where M-phase can form, but fails to cover the surface before the S-phase begins to appear, the S-phase was observed to form only on bare regions of the HOPG (Fig. 5a-d, Supplementary Fig. 8, Supplementary Video 1). That is, we observed no instances in which S-phase islands appeared either on top of existing M-phase islands or through direct transformation from the M- to the S-phase. Moreover, while the M-phase grew row-by-row via classical monomer-by-monomer addition, the S-phase followed an aggregation-based growth pathway analogous to oriented attachment, which involved individual row nucleation followed by a process of row diffusion, aggregation and ordering[26–28] (Fig. 3, Fig. 5e-p, Supplementary Fig. 9, Supplementary Video 1). Because the S-phase rows exhibited high mobility compared to the M-phase, once a row nucleated, subsequent diffusion on the HOPG surface led to aggregation with nearby rows, followed by rearrangement to the final S-phase structure (Fig. 3, Fig. 5e-p, Supplementary Video 1). As a consequence, the degree of order of the S-phase islands increased over time as peptide rows stabilized one another and aligned along the preferred lattice directions. However, the orientations of the S-phase islands on the graphite lattice were distinct from those of the M-phase, exhibiting a difference of either ~ -18° or ~23° between the sets of orientations, with that at ~23° being dominant (Supplementary Fig. 5).

In contrast to the situation at 4 µM peptide concentration where M-phase coverage is incomplete, at 5 µM M-phase coverage is complete before the S-phase begins to appear. Thus the S-phase rows are created within the confined spaces at boundaries between M-phase islands or at defect sites within the islands and, hence, lose their ability to rotate and diffuse (Fig. 6). As a result, the M-phase islands serve as templates that force the S-phase rows to align along an otherwise unfavorable set of directions — namely the armchair direction of HOPG (Fig. 6i-l, Supplementary Fig. 10). Combined with the two preferred sets of directions adopted by freely nucleating and diffusing S-phase islands, this extra set of directions templated by M-phase islands leads to the observed three direction sets seen for S-phase islands after long incubation times (Fig. 2a, Supplementary Fig. 11). In this way, the pattern of the M-phase is imprinted upon the subsequent pattern of the S-phase. The existence of a dominant and a secondary direction for S-phase islands combined with this concentration-dependent phenomenon of M-phase templating of S-phase islands, leads to the complex mix of geometric relationships observed at arbitrary timepoints (See Supplementary Text and Supplementary Fig. 10, Supplementary Fig. 11 for details).

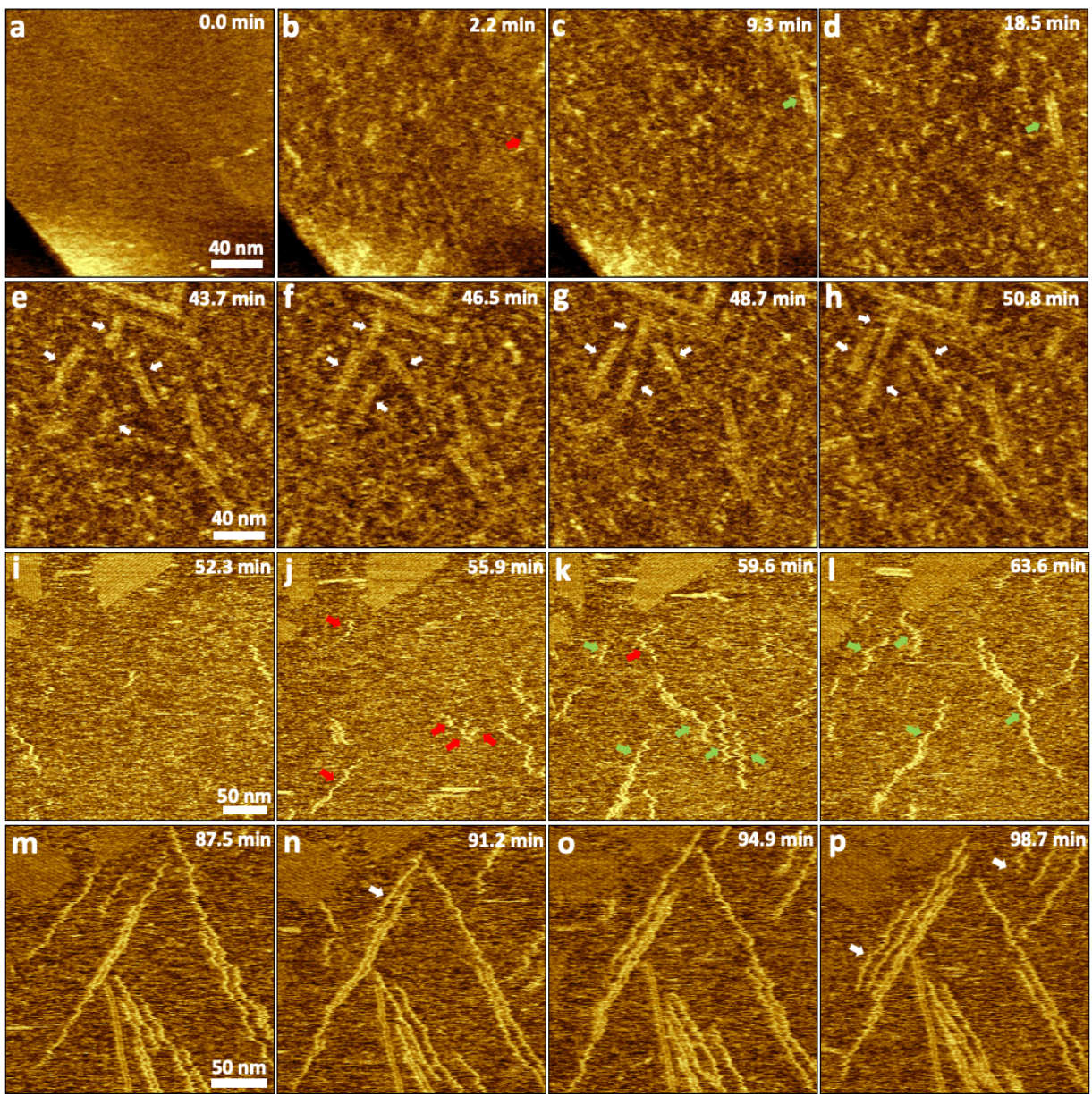

**Fig. 5| Nucleation and growth mechanism of the S-phase on HOPG. a-h**, Nucleation and growth of S-phase islands on bare HOPG when 4 μM solution was flowed through the AFM liquid cell, which initially contained only pure water. The actual concentration is expected to be

lower than 4 µM. **i-p**, S-phase formation at 4 µM, a concentration at which S-phase islands begin to form in bare areas of the HOPG before the M-phase can cover the surface, following a growth pathway that involves row-by-row nucleation, diffusion, aggregation and ordering. (The waviness of the rows is observed because the rows are mobile on the surface.) White arrows indicate diffusion of a row. Red and green arrows mark the formation of new rows and the elongation of existing rows, respectively.

**The key role of fluctuations due to reversible peptide binding**

Given that S-phase islands can only form on bare regions of the HOPG surface, the question arises as to how they are able to form and eliminate the M-phase at high peptide concentration (5 µM) (Fig. 6, Supplementary Fig. 12, Supplementary Fig. 13, Supplementary Fig. 14) when M-phase islands form first and rapidly grow (within ~10 min) to fully cover the surface (Supplementary Fig. 15) before any S-phase islands appear? This transformation is possible, because the formation of M-phase islands along the three equivalent directions of the graphite lattice creates defects along the boundaries between the islands. In addition, some defects exist within the islands. The continuous attachment and detachment of peptides at the ends of the rows at these inter-island boundaries and intra-island defects lead to fluctuations in local coverage (Supplementary Fig. 16) that create transient windows of opportunity for nucleation of S-phase islands (Fig. 6i-l). Once nucleated, the S-phase then grows as the metastable M-phase islands dissolve, creating further available space for more S-phase nucleation and growth (Fig. 6).

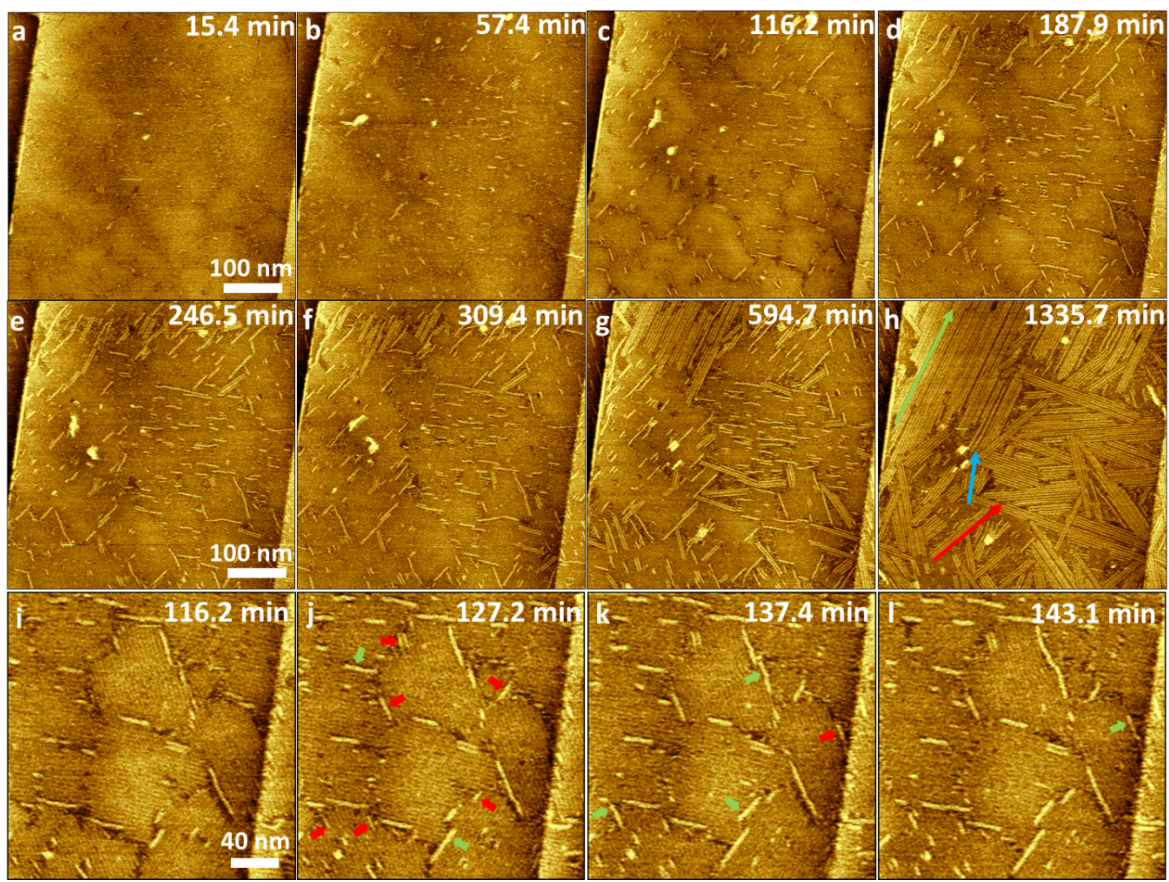

**Fig. 6| Snapshots showing the developments of the S-phase at 5 µM MoSBP1 due to fluctuations at M-phase island edges. a-h**, Even when M-phase islands rapidly and fully cover the surface before any S-phase islands can form, the S-phase appears gradually and comes to dominate over time. Arrows mark 3 different sets of directions of S-phase peptide rows. **i-l**, zoomed-in images show creation of the S-phase occurring exclusively at inter-island boundaries and intra-island defects of the M-phase. Red and green arrows mark the formation of new rows and the elongation of existing rows, respectively.

**Discussion and Conclusions**

The results presented above show that MoSBP1 forms two distinct ordered phases on HOPG surfaces and that these phases can coexist for extended time periods. However, the metastable phase, which is kinetically preferred and assembles rapidly on HOPG following a row-by-row growth pathway, can only form at high peptide supersaturation. In contrast, the thermodynamically stable phase forms more slowly via row-by-row nucleation followed by diffusion, aggregation and ordering. The metastable phase dissolves as the stable one grows and serves as a reservoir of peptide monomers. Thus, even when the M-phase is initially the only one present the S-phase comes to dominate over time.

The M-phase exhibits a smaller repeating unit (~3.9 nm × 1.9 nm) and a strict epitaxial relationship with the underlying HOPG substrate, aligned along three equivalent armchair directions. In addition, there is no sign of row mobility; rows are straight and do not diffuse on the surface. In contrast, the S-phase has a larger repeating unit (~4.4 nm × 1.1 nm), closer to that observed on $MoS_2$ (~4.7 nm × 1.1 nm), and shows a weaker epitaxial relationship with HOPG, with three sets of equivalent directions, as well as wavy row morphologies and diffusion across the surface due to significant mobility.

Based on these observations, it is reasonable to hypothesize that the M-phase is strongly bound to HOPG. However, the smaller lattice of HOPG likely compresses the MoSBP1 peptide dimers into a reduced unit cell (~3.9 nm × 1.9 nm), introducing strain within the dimers and at the peptide–substrate interface. This strain renders the M-phase thermodynamically less favorable. In contrast, the S-phase is loosely bound to HOPG, leading to the significant mobility and multiple permissible orientations. While the S-phase maintains a dimer structure closer to what is presumably the thermodynamically preferred configuration observed on $MoS_2$, the reduced strain leads to greater intrinsic stability despite weaker substrate binding.

The results also demonstrate that there is no direct transition from the metastable to the stable phase. Rather, S-phase islands nucleate and grow only on bare HOPG within the gaps between M-phase islands or between rows within an island, demonstrating that S-phase rows are created through independent nucleation events directly from monomers in solution and not through conformational changes or realignment of M-phase islands. Consequently, when the M-phase is formed at very high peptide concentrations so that it completely covers the surface, the S-phase can only form in the transient spaces created through fluctuations of M-phase island edges arising from the reversibility of peptide binding.

The observed sequence of events also illustrates Ostwald's Rule of Stages occurring through the dissolution-reprecipitation process common to 3D crystal systems. However, despite the qualitative agreement, significant modifications to the theoretical rationale are needed, because the nuclei for these 2D crystals are 1D structures for which there is no free energy barrier[15]. Consequently, the occurrence of Ostwald's rule of stages, originally framed in terms of the thermodynamic control on nucleation, namely the magnitude of the free energy barrier, must arise here for purely kinetic reasons. To understand the difference, note that there are two energy barriers that control nucleation rates. For 3D particles and 2D islands, there is a free energy barrier that arises for purely thermodynamic reasons due to the competition between the increase in surface energy and the decrease in bulk free energy. In contrast, for systems of 1D rows, there is no free energy barrier. However, kinetic energy barriers still exist, associated with atomistic processes such as folding/unfolding, desolvation, and other structural rearrangements. Moreover, in the case of row formation on a surface the lifetime of the first dimer pair that must initiate the row — and hence the binding energy to the substrate — directly effects the kinetics or row formation. Thus, which phase nucleates faster depends on these kinetic factors, while the stability of the phases depends on the differences in free energy between the dissolved state and the assembled state on the surface.

As shown in the previous study[15] on $MoS_2$, the nucleation rate $J_n$ of a row can be written as: $J_n = K(c - c_e)$, where $K$, $c$, $c_e$ are the rate constant, peptide concentration, and peptide solubility, respectively. As shown in that study, the rate-limiting step that determines the magnitude of $K$ is the formation of a dimer, which is the first stable unit of a row. Given that $c_e$ is smaller for the S-phase, the term $(c - c_e)$ will be larger and, thus, its slower nucleation rate at high peptide concentrations ($J_{n,S} < J_{n,M}$) implies that $K_S < K_M \cdot [(c - c_{e,M})/(c - c_{e,S})] < K_M$ where the subscripts denote the S- or M-phase. Thus, writing $K$ as $K_0 \exp(-\Delta E_a/kT)$, where $\Delta E_a$ is the activation energy barrier for the creation of a dimer, the results presented above imply that the order of the appearance of phases is a consequence of the magnitude of $\Delta E_a$, which now replaces free energy barrier $\Delta G_c$ in determining that order[29]. In the case of MoSBP1 crystals on HOPG, evidently $\Delta E_a$ is larger for the S-phase since $K_S < K_M$.

The reason for the higher values of $\Delta E_a$ for the more stable phase can be understood from both specific analyses of MoSBP1 binding to the surface, as well as general considerations of chemical reactivity. Assembly of MoSBP1 on the surface requires a conformational change from a coiled state in bulk solution to the extended state adopted in the 2D crystal, as well as the displacement of adsorbed water[15]. In the case of $MoS_2$, molecular dynamics simulations predict that the adsorption free energy for MoSBP1 predominantly originates from the replacement of weakly bound water molecules in direct contact with the hydrophobic surface; these waters gain more hydrogen bonds and increased entropy upon release into the solution[15]. Formation of a building unit for the S-phase on the surface, which is stretched to a unit length of 4.4 nm compared to only 3.9 nm for M-phase, should be associated with the breaking of more intramolecular interactions during extension from the coiled state, as well as the removal of more water molecules from the surface, leading to a higher activation barrier, and thus a smaller rate constant $K$, for transitioning from the solution state to the bound state.

On the other hand, following the general wisdom of chemical kinetics that a smaller bond energy results in a lower kinetic barrier to form or rupture the bond in a chemical reaction[30], we can also expect that the activation barrier for attachment of a molecule to create a dimer of any 1D (or quasi-1D) phase will exhibit a trend towards larger values with increasing phase stability (i.e., decreasing solubility). Thus, the rate constant $K$ should typically be larger for the metastable phase, consistent with the above discussion and our observations. As with the correlation between surface energy and phase stability that underlies the thermodynamic rationale for Ostwald's Rule in 2D and 3D, while this kinetic rationale is expected to hold in general, it is unlikely to apply universally.

The kinetic version of Ostwald's Rule leads to a specific prediction for the concentration at which a crossover occurs from a regime where the M-phase appears first to one in which it is rarely seen. The ratio of nucleation rates $J_i$ between the M- and S- phases is given by:

$$\frac{J_M}{J_S} = \frac{K_M(c-c_{e,M})}{K_S(c-c_{e,S})} \qquad (1)$$

Because $c - c_{e,M} < c - c_{e,S}$ and $K_M > K_S$, in the regime where peptide concentration $c$ is large, $(c - c_{e,M})/(c - c_{e,S})$ gets close to 1 and $K_M/K_S$ dominates, which leads to the rapid formation of M-phase ($J_{n,M}/J_{n,S} > 1$). While at small enough concentrations, where $(c - c_{e,M})/(c - c_{e,S})$ takes over, the first phase to appear will tend to be the S-phase. The crossover point occurs for:

$$c_{crossover} = \frac{K_M c_{e,M} - K_S c_{e,S}}{K_M - K_S} = c_{e,M} + \frac{K_S(c_{e,M} - c_{e,S})}{K_M - K_S} > c_{e,M} \qquad (2)$$

When $c_{e,S} < c < c_{e,M}$, the M-phase is not stable and the S-phase is the only one that can nucleate. When $c_{e,M} < c < c_{crossover}$, the S-phase still nucleates faster and its formation will lower the concentration and further decrease $J_{n,M}$. Thus, nucleation of the M-phase is rare. This regime is best represented by the 1 µM case (Fig. 1a) where only a wavy S-phase and no M-phase can be observed. It can also be represented by the 4 µM case at later times (Fig. 3, 40 to 60 min) when M-phase islands are still growing ($c_{e,M} < c$), but nucleation of new M-phase islands is rare compared to that of the S-phase ($c < c_{crossover}$). When $c > c_{crossover}$, the M-phase nucleates rapidly and dominates the surface at early times. In this experimental system, the $c_{crossover}$ should be a number between 1 µM and 4 µM (1 µM $< c_{crossover} <$ 4 µM). This reversal of phase preference with increasing concentration is similar to what is expected with increasing supersaturation in 2D and 3D systems that exhibit Ostwald's Rule of Stages.

These findings also highlight the crucial role of M-phase fluctuations in enabling S-phase formation; when the supersaturation is high enough to completely cover the HOPG surface with M-phase islands, the S-phase can only nucleate at inter-island boundaries or intra-island defects where the fluctuations at the ends of the rows transiently expose bare graphite where the S-phase can nucleate. Moreover, through comparison to results obtained previously on MoS$_2$, on which a single, immobile phase forms, these findings demonstrate the ability of the substrate to select both the structures and assembly pathways of stable and metastable phases.

The findings reported here help to establish a framework for interpreting self-assembly of molecules on inorganic interfaces that is rooted in well-developed classical theories and can be used to guide design of both molecules and substrates in order to achieve assembly of targeted 2D hybrid materials[31–33]. Moreover, the resulting insight into the use of concentration to trigger Ostwald's Rule of Stages suggests a strategy for on-surface synthesis aimed at creating new organic materials[34–36] by capturing transient 1D and 2D phases that present otherwise inaccessible arrangements of organic molecules at interfaces.

Understanding Ostwald's rule in 1D is crucial for controlling length distributions and phase stability in supramolecular polymers, peptide nanofibers, and amyloid fibrils, with important implications for biomaterials design, neurodegenerative disease research, organic electronics, and interfacial crystal engineering. Thus the findings reported here have the potential to impact fields ranging from soft matter physics and chemical biology to nanomaterials synthesis and functional device engineering.[37–41]

## Methods

**Preparation of peptide stock solution.** Lyophilized peptides (0.001 g) were mixed with 20 mL nuclease-free water (Ambion, USA) in a centrifuge tube and ultra-sonication was used to facilitate dissolution. The final concentrations of peptide stock solutions were around 0.06 mM. Solutions were diluted to 1 - 5 µM for the AFM experiments.

**AFM imaging.** 40 µL of peptide solution was added on top of a freshly cleaved HOPG surface (1 cm × 1 cm, Ted Pella) for incubation inside the AFM at room temperature. AFM under flowing conditions was performed in a small droplet of solution with a volume ranging from 50 to 100 µL using a perfusion cantilever holder equipped with PTFE tubing. Solution exchange was performed via a syringe pump, and a minimum of 2 mL of fresh solution was flowed between each imaging condition to guarantee complete replacement. In situ images were captured as described previously[9] using silicon probes (SNL, k: 0.12 N/m or 0.24 N/m; tip radius: 2 nm; Bruker) and silicon nitride probes (OTR4 and OTR8, k: 0.08 N/m, 0.15 N/m, tip radius: 15 nm; Bruker) under tapping mode with a Cypher ES AFM (Asylum Research) at room temperature. Images were analyzed using Gwyddion SPM data analysis software.

**Data availability**

The authors declare that the data supporting the findings of this study are available within the paper and its supplementary information files.

**Acknowledgements**

Peptide synthesis and AFM studies were supported by the NSF EFRI 2DARE Program (NSF EFRI-1433541). High-resolution AFM imaging and AFM imaging under flowing conditions were supported by the U.S. Department of Energy (DOE), Office of Basic Energy Sciences (BES) under contract FWP 72448 as part of the Energy Frontier Research Center program: CSSAS – The Center for the Science of Synthesis Across Scales, located at the University of Washington under Award Number DE-SC0019288. Development of physical models was supported by the US DOE BES Office of Basic Energy Sciences, Division of Materials Science and Engineering at Pacific Northwest National Laboratory (PNNL) under award FWP 65357. In situ AFM was performed at the PNNL. PNNL is a multi-program national laboratory operated by Battelle for the U.S. Department of Energy (DOE) under Contract DE-AC05-76RL01830. Y.

Huang acknowledges support from the Office of Naval Research (ONR), grant number N00014-18-1-2491. We thank E. Zhu for providing peptide samples.

**Author contributions**

J.C. and Y.X. performed the AFM experiments and data analysis, and wrote the manuscript; M.Z. performed the AFM experiments and data analysis; Y.H. designed the study; J.J.D.Y. designed the study, performed data analysis, and wrote the manuscript.

**Competing interests**

The authors declare no competing interests.

**Additional information**

Supplementary information is available in the online version of the paper.

Supplementary Information

**Ostwald's Rule of Stages in one-dimension**

Jiajun Chen[1,2,†], Ying Xia[1,2,†], Mingyi Zhang[2,3], Yu Huang[4,5], and James J. De Yoreo[1,2,*]

[1]Department of Materials Science and Engineering, University of Washington, Seattle, WA 98195, USA

[2]Physical Sciences Division, Pacific Northwest National Laboratory, Richland, WA 99352, USA

[3]School of Aerospace and Mechanical Engineering, University of Oklahoma, Norman, OK 73069, USA

[4]Department of Materials Science and Engineering, University of California, Los Angeles, CA 90095, USA

[5]California NanoSystems Institute, University of California, Los Angeles, CA 90095, USA.

†Co-first authors

*Corresponding author. Email: james.deyoreo@pnnl.gov

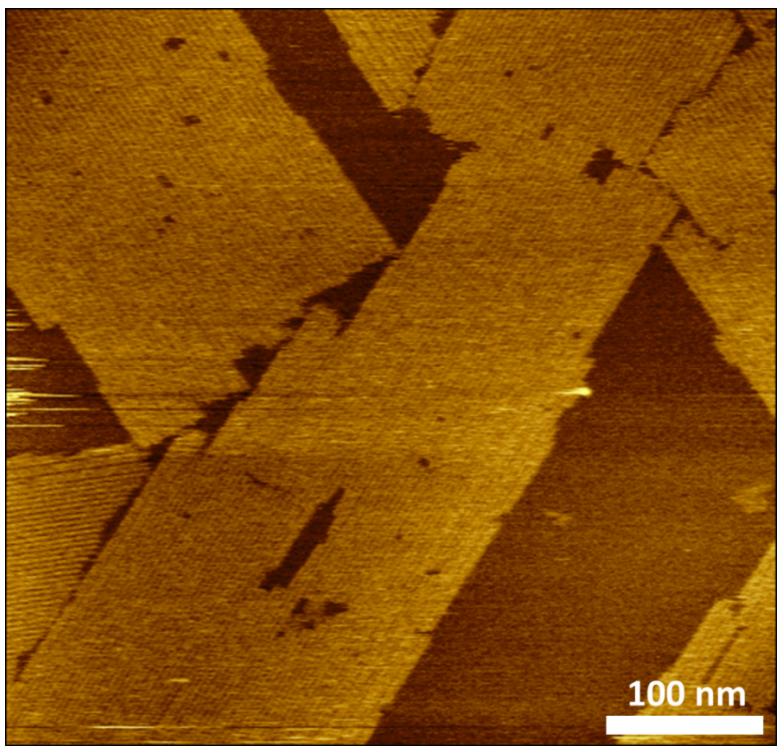

**Supplementary Figure 1| AFM image of the self-assembled patterns of MoSBP1(0.75μM) on MoS$_2$.** The 2D islands consisted of parallel rows and aligned along 3 equivalent directions.

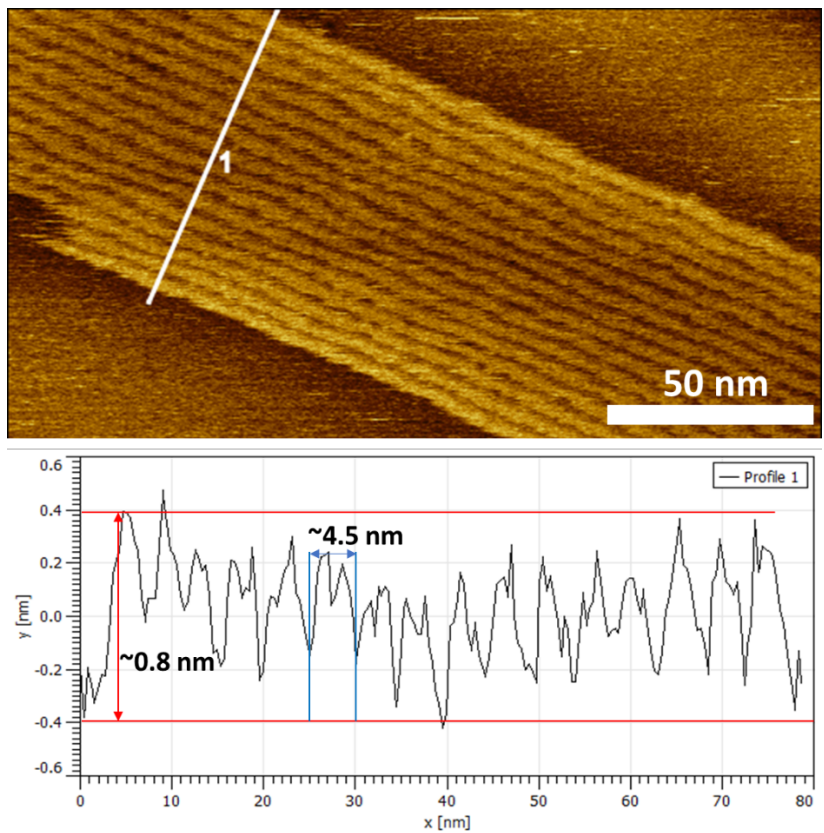

**Supplementary Figure 2| High-resolution AFM image and height profile of the self-assembled patterns of MoSBP1 (S-phase) on HOPG.** The 2D island consists of parallel rows with uniform spacing. The height of the island is ~0.8 nm, and the center-to-center distance of rows is ~4.5 nm.

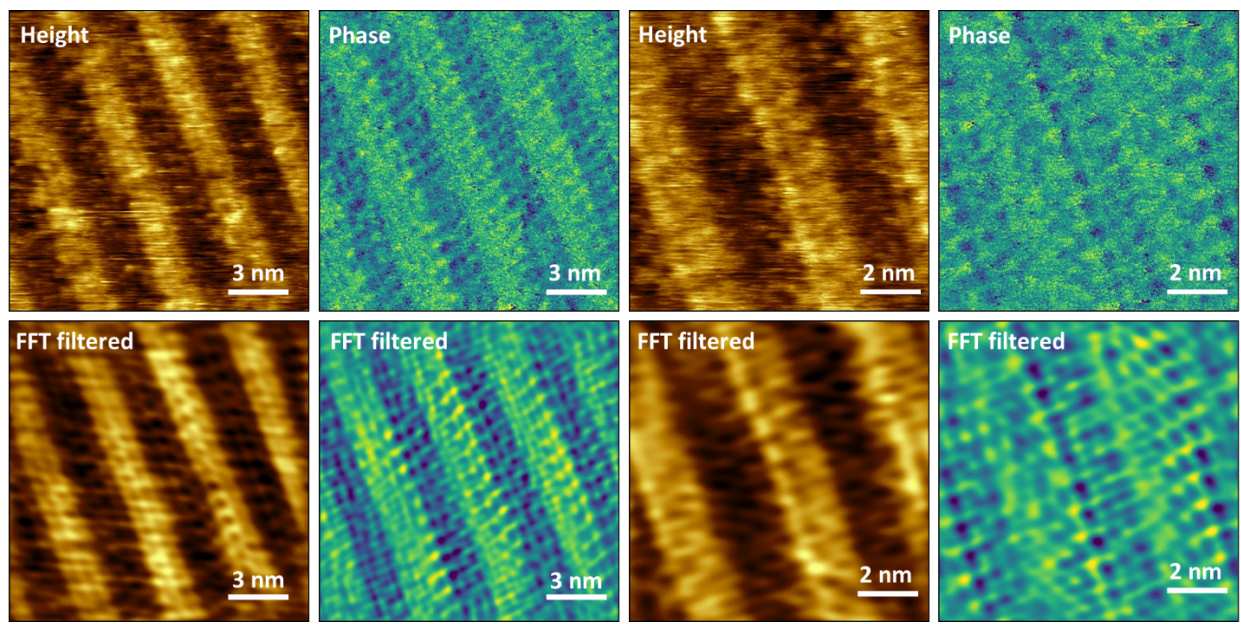

**Supplementary Figure 3| High-resolution AFM height and phase images of the self-assembled patterns of MoSBP1 (S-phase) on HOPG.** The top row shows the raw images, while the bottom row displays the FFT-filtered images.

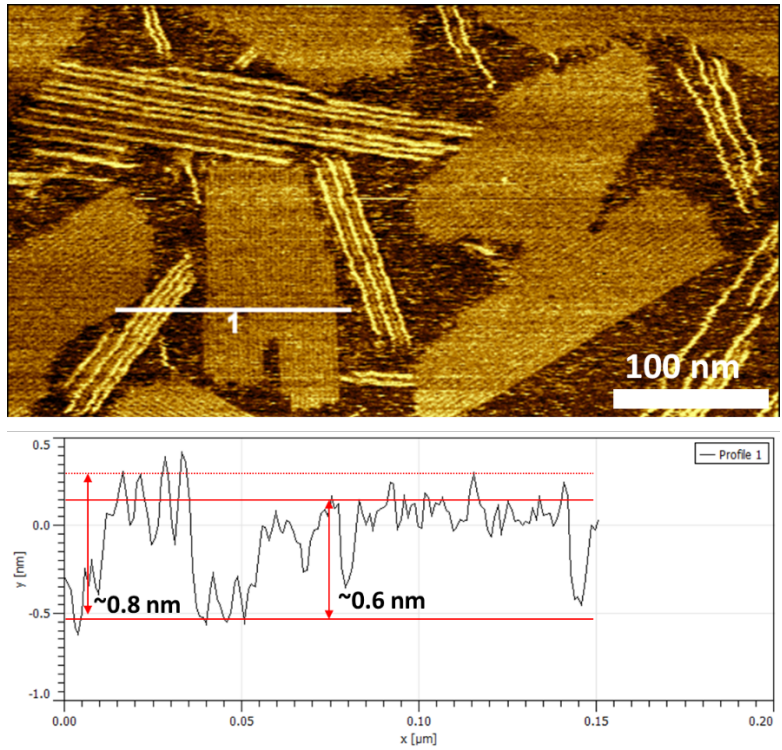

**Supplementary Figure 4| Coexistence of two different phases on graphite surfaces**. AFM image and height profile of the M-phase and S-phase structures.

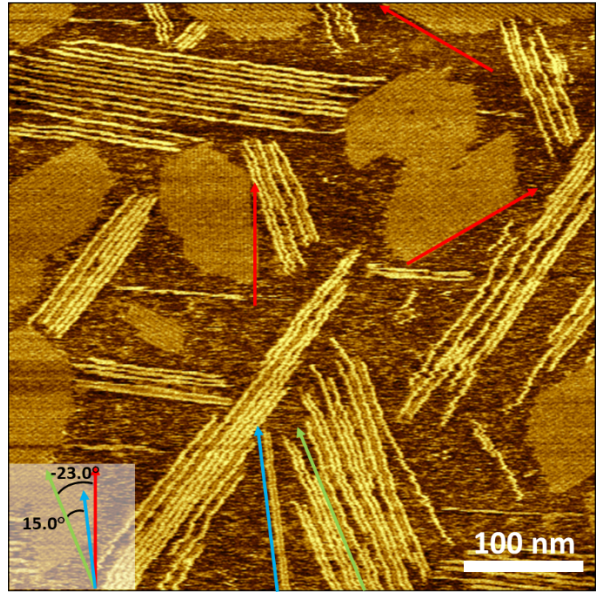

**Supplementary Figure 5| Preferred directions of M-phase and S-phase assembled at 4 µM.** The M-phase islands only aligned along three equivalent directions (one set of directions, red arrows). The preferred orientation of the S-phase (green arrow) was -23° different from that of the M-phase (red arrow). The minority orientation of the S-phase (blue arrow) was 15° different from the preferred orientation of the S-phase (green arrow).

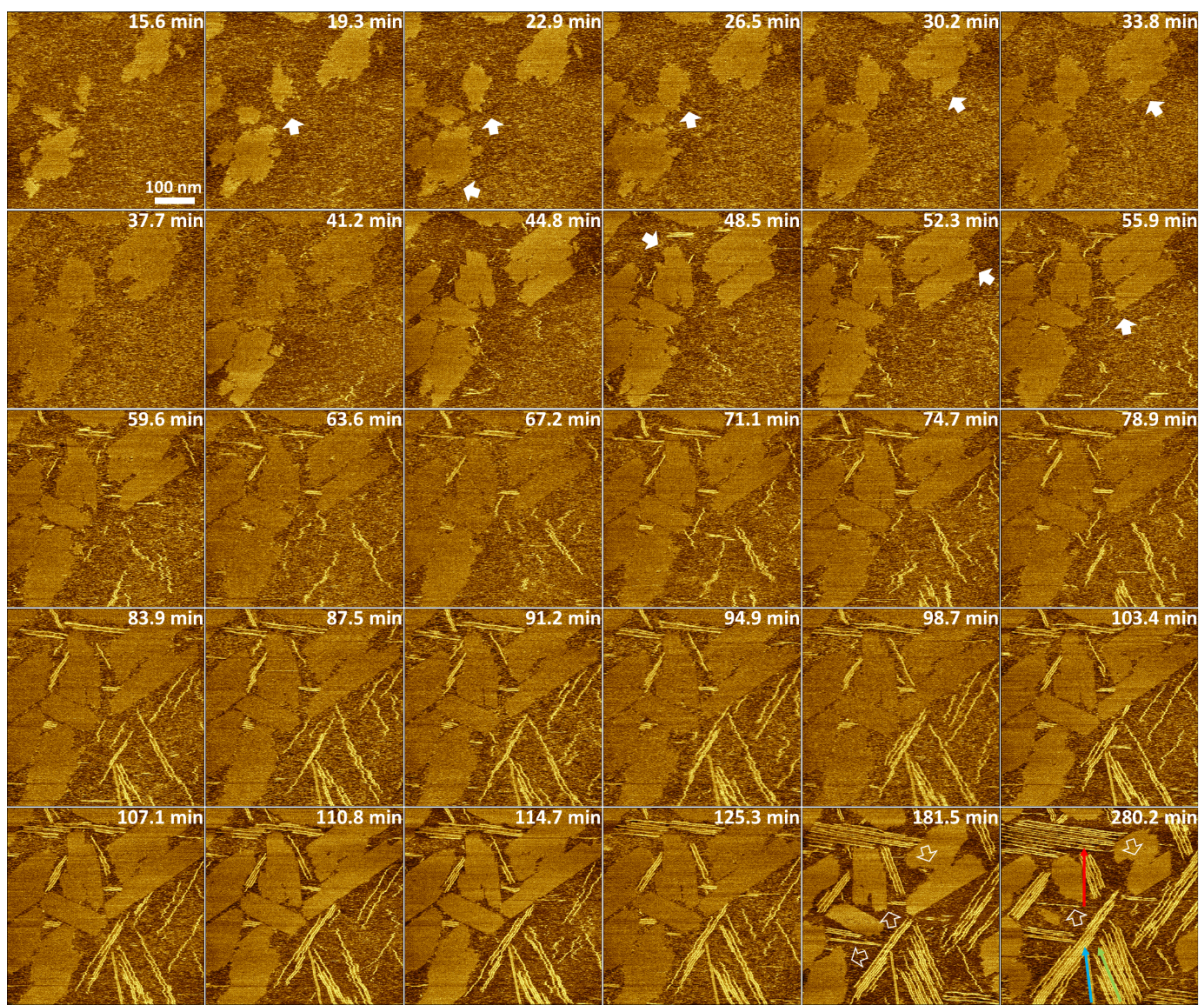

**Supplementary Figure 6| Developments of the M-phase and S-phase at 4 µM MoSBP1 on HOPG.** M-phase islands were created first and then gradually dissolved. S-phase formed slowly and started to dominate over time. Solid and open white arrows mark the growth and dissolution of M-phase, respectively. Colored arrows in final panel denote the 3 different sets of directions of the peptide rows.

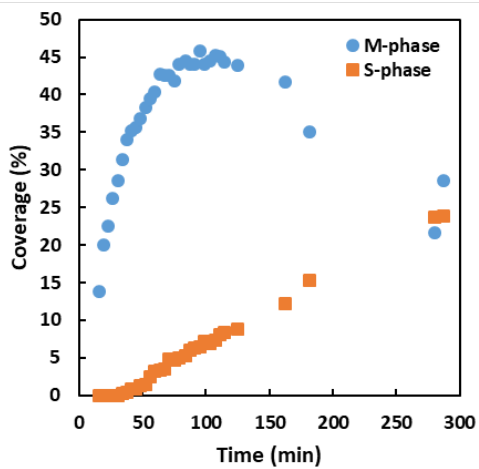

**Supplementary Figure 7| Coverages of M-phase and S-phase in Supplementary Figure 6 vs. time.** M-phase formed much faster but then gradually dissolved as the S-phase grew and consumed free monomers in the solution. S-phase formed slowly and started to dominate over time.

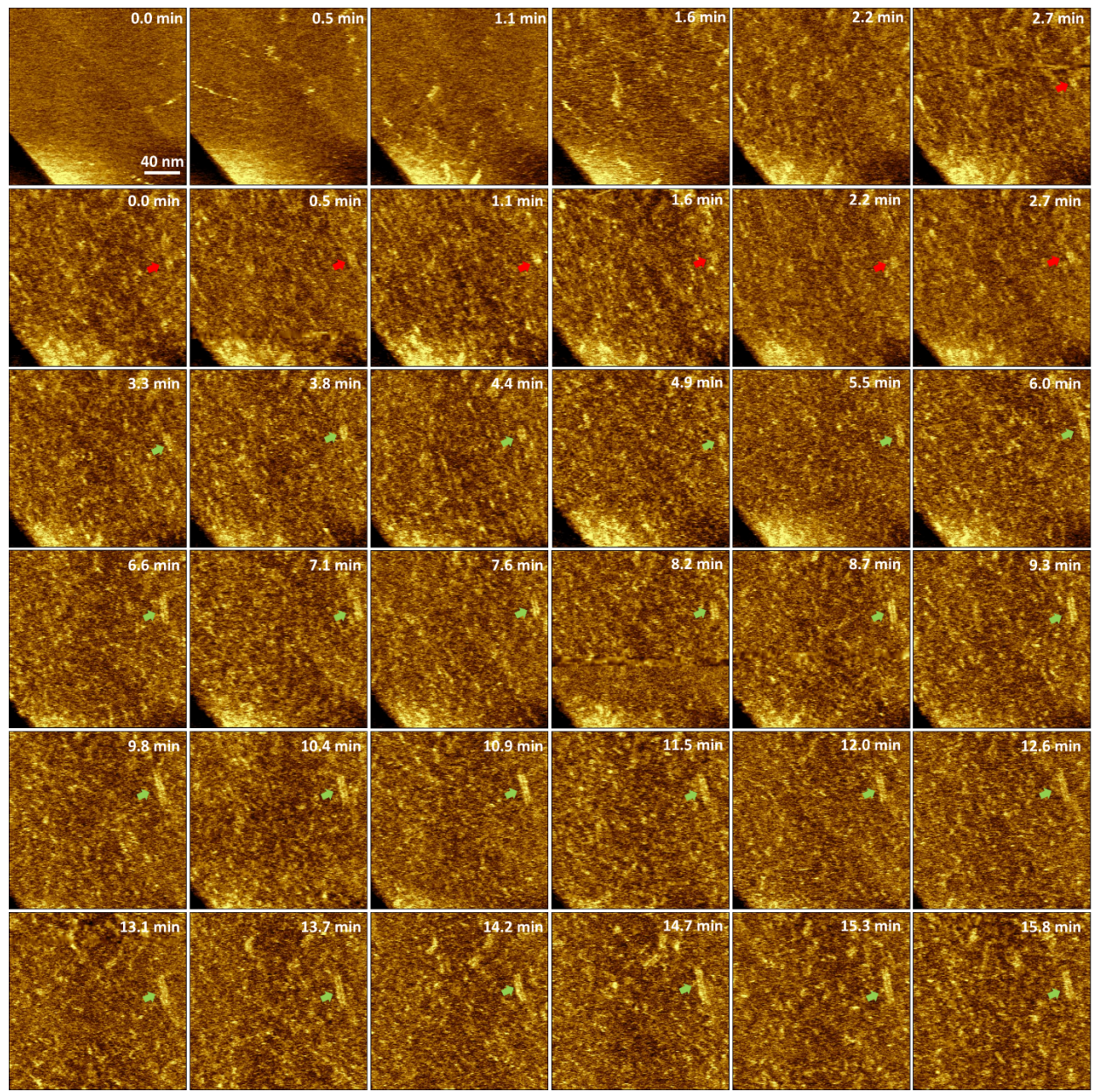

**Supplementary Figure 8| *In-situ* AFM images of the nucleation of the S-phase of MoSBP1 (<4 μM) on HOPG.** Rows of S-phase nucleate on the HOPG surface after we flow in 4 μM peptide solution to the AFM cell initially containing only pure water. Red and green arrows mark the formation of new rows and the elongation of existing rows, respectively.

The population of the S-phase is low and was not counted in the early stage, which results in a nonzero y-intercept for the S-phase in Supplementary Fig. 7. However, the high-speed data in Supplementary Fig. 8 show that S-phase nucleation can be observed as early as 2 minutes after peptide solution injection.

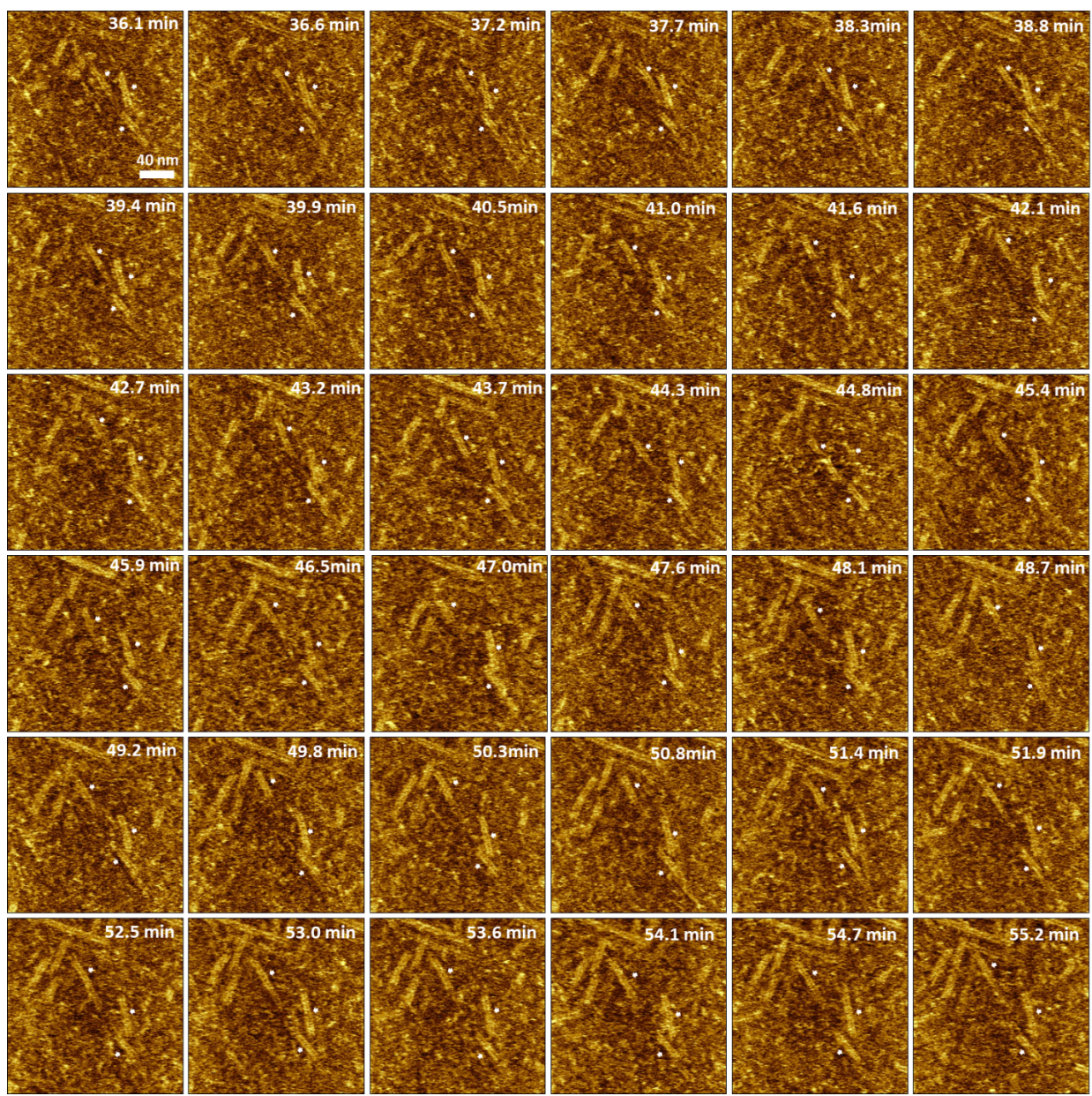

**Supplementary Figure 9|** *In-situ* **AFM images showing the diffusion, aggregation and ordering of rows of the S-phase of MoSBP1 on HOPG after flowing in 5 μM peptide solution to the AFM cell initially containing pure water .** The identical rows are marked with arrows for clearer illustration. The results demonstrate the high mobility of the S-phase rows, attributed to the weak interaction between the peptide and the substrate.

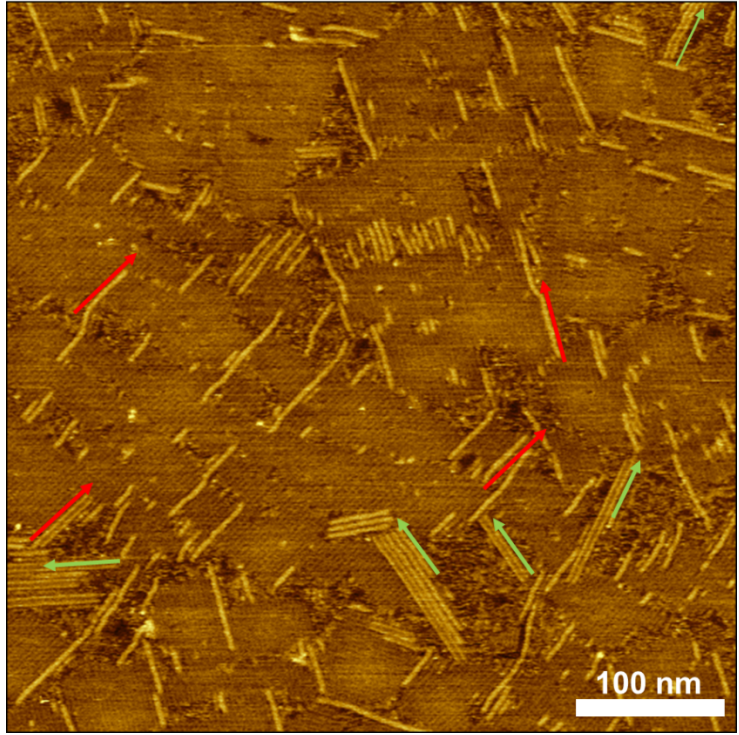

**Supplementary Figure 10| Rows comprising the S-phase were sometimes templated to align along the orientations of M-phase islands.** S-phase rows created at larger gaps could rotate and align along preferred directions (marked with green arrows). S-phase rows that formed at the boundaries or defect sites with limited space were forced to align along the directions of M-phase islands (marked with red arrows).

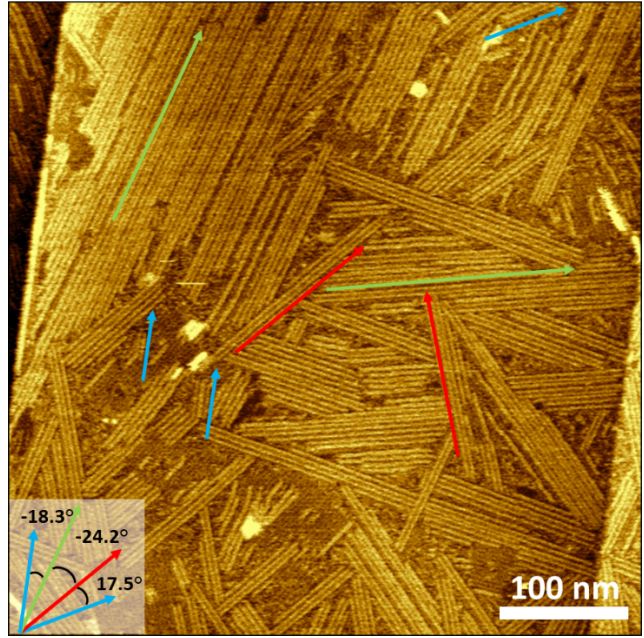

**Supplementary Figure 11| Directions of S-phase islands assembled at 5 μM shown in Fig. 5.**
S-phase domains along the directions marked with red arrows were templated by previous M-phase islands. In this way, the pattern of the M-phase was imprinted upon the subsequent pattern of the S-phase. The preferred orientation of the S-phase (green arrow) was -24.2° different from that of the M-phase (red arrow). The minority orientation of the S-phase (blue arrow) was -18.3° different from the preferred orientation of the S-phase (green arrow).

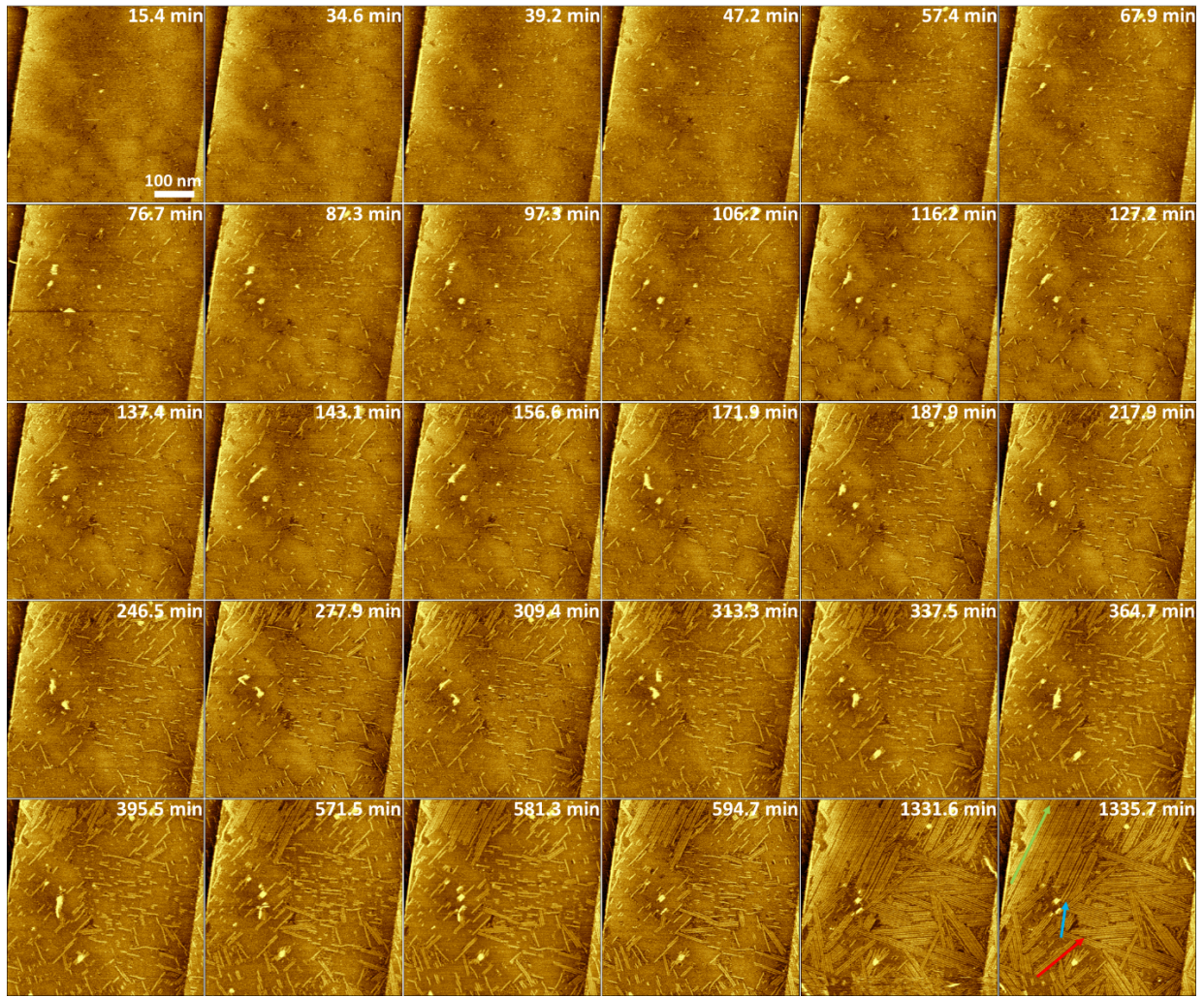

**Supplementary Figure 12| Developments of the S-phase at 5 µM MoSBP1 due to fluctuations at M-phase island edges.** M-phase islands fully covered the surface rapidly before any S-phase islands could form. The S-phase appeared gradually and still came to dominate over time. Arrows mark 3 different sets of directions of peptide rows.

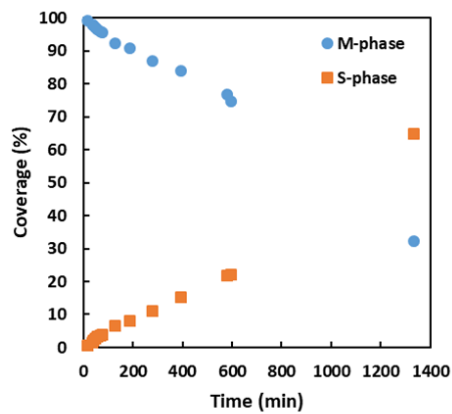

**Supplementary Figure 13| Coverages of M-phase and S-phase in Supplementary Figure 12 vs. time.** M-phase formed rapidly and fully covered the surface at the early stage, and then gradually dissolved as the S-phase grew and consumed free monomers in the solution. S-phase formed slowly and started to dominate over time.

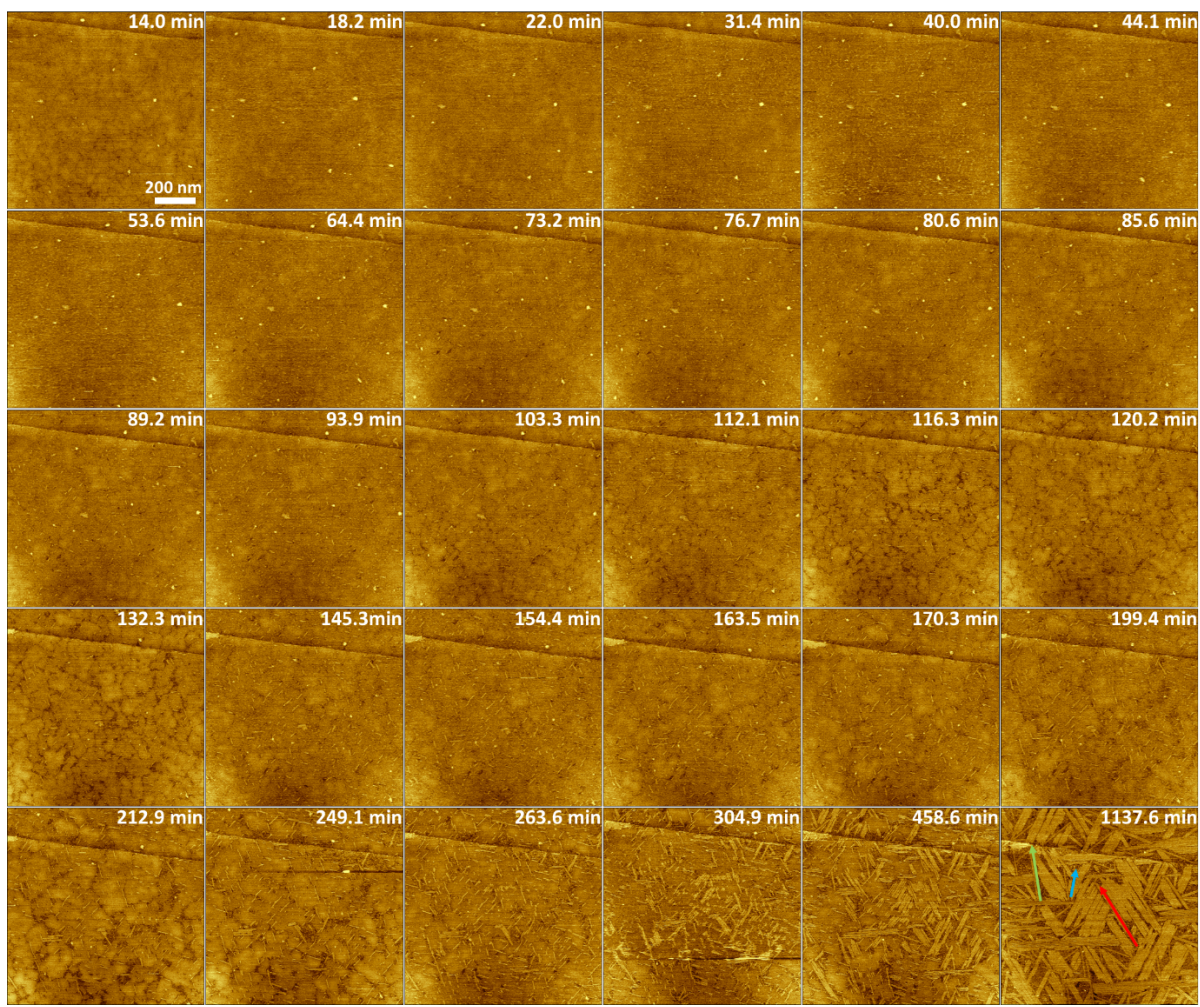

**Supplementary Figure 14| Development of S phase islands following initial coverage by M-phase islands at 5 μM on HOPG.** Arrows mark 3 different sets of directions of peptide rows in the S-phase.

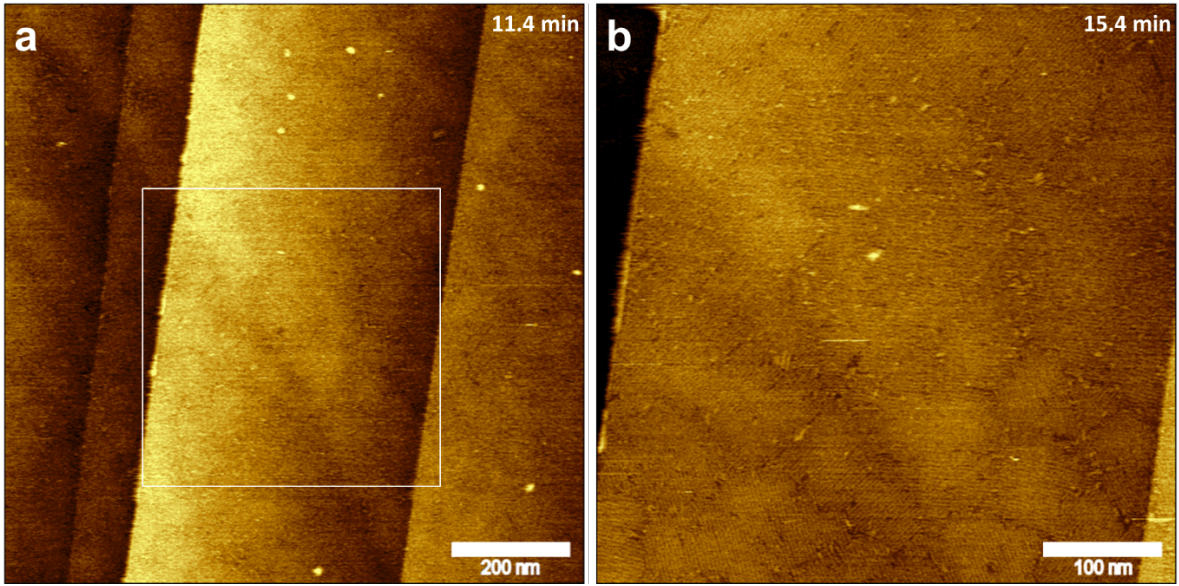

**Supplementary Figure 15| In situ AFM images of assembled structures of MoSBP1 (5 µM) at an early time on a graphite surface. a**, the graphite surface was fully covered by a layer of peptides. **b**, High-resolution image of the selected part in (a), showing that the surface was covered by ordered domains.

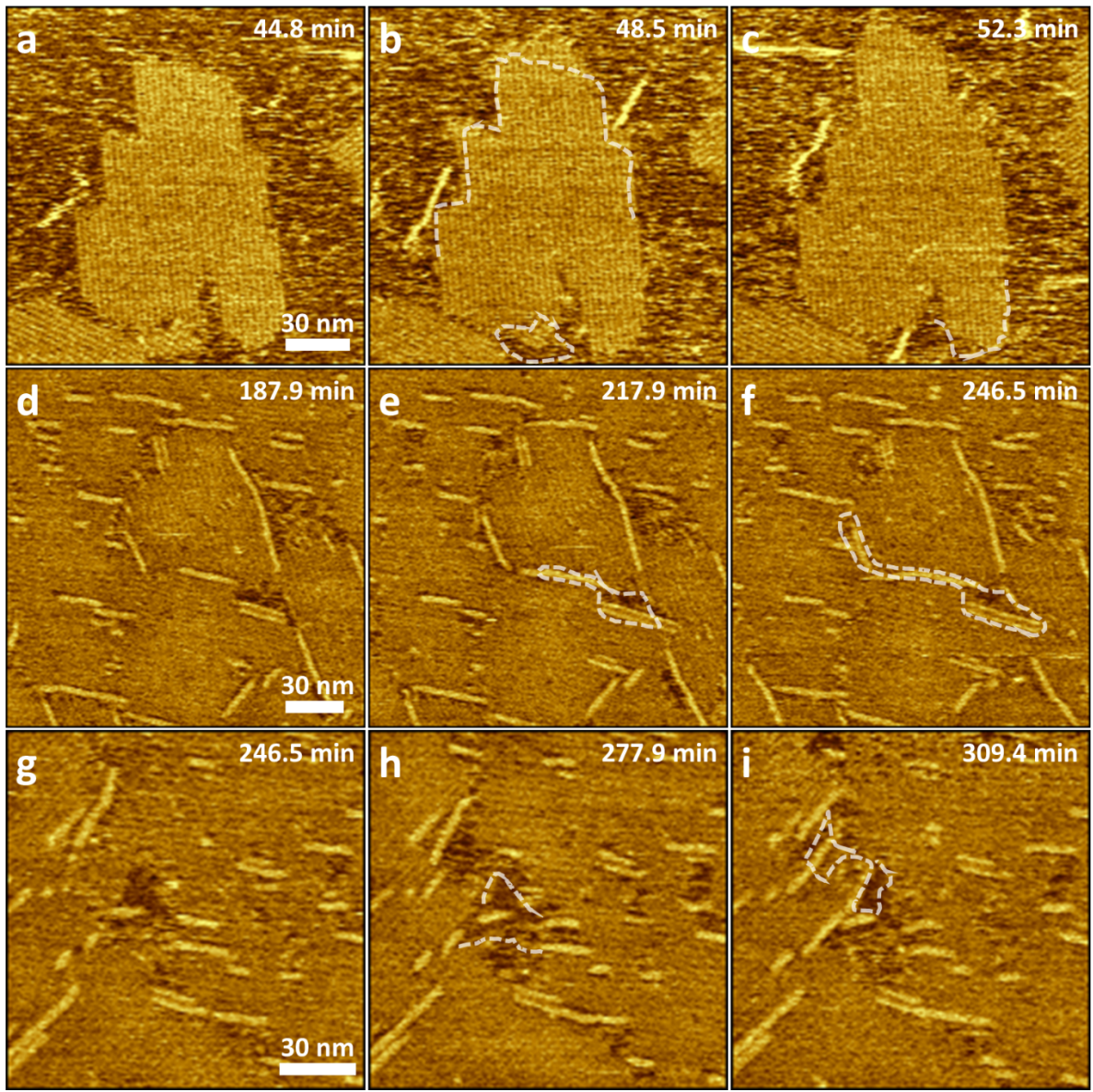

**Supplementary Figure 16| Fluctuations in local coverage of M-phase islands.** The reversible binding of peptides at the ends of the rows at these inter-island boundaries and intra-island defects led to fluctuations in local coverage that created transient windows of opportunity for the nucleation of S-phase islands. Dashed lines mark the previous locations of M-phase boundaries.

**Supplementary Text**

<u>Further explanation about the geometric relationships of S-phase and M-phase</u>

The preferred orientation of the S-phase (green arrow) was around ±24° different from that of the M-phase (red arrow). The minority orientation of the S-phase (blue arrow) was around ±18° different from the preferred orientation of the S-phase (green arrow) (Fig. 2a, Supplementary Fig. 14). There was also a small possibility for the combination of -24° and 18° which lead to a small angle difference (~6°) between the preferred orientation of the M-phase (red arrow) and the minority orientation of the S-phase (blue arrow) as shown in Supplementary Fig. 5 instead of ±18° (Fig. 2a, Supplementary Fig. 11). It is also worth noting that the final major direction of S-phase in the case of 5 µM where nucleation, rotation, and diffusion of S-phase peptide rows are restricted can be either the preferred orientation of the S-phase (green arrow) (Supplementary Fig. 11) or the preferred orientation of the M-phase (red arrow) (Fig. 2a), since the growth of these S-phase domains could be highly dependent on their local environments and the population of templated S-phase was significantly increased (Supplementary Fig. 10).